\documentclass[10pt, conference, compsocconf]{IEEEtran}

\ifCLASSINFOpdf
\usepackage[pdftex]{graphicx}
\else
\fi
\usepackage{subcaption}

\usepackage{url}

\usepackage{flushend}

\begin{document}
%
\title{Automatically Identifying Fake News in Popular Twitter Threads}

\author{\IEEEauthorblockN{Cody Buntain}
\IEEEauthorblockA{Intelligence Community Postdoctoral Fellow\\
University of Maryland, College Park\\
Email: cbuntain@cs.umd.edu}
\and
\IEEEauthorblockN{Jennifer Golbeck}
\IEEEauthorblockA{College of Information Studies\\
University of Maryland, College Park\\
Email: golbeck@cs.umd.edu}
}


%


\maketitle

\begin{abstract}
Information quality in social media is an increasingly important issue, but web-scale data hinders experts' ability to assess and correct much of the inaccurate content, or ``fake news,'' present in these platforms.
This paper develops a method for automating fake news detection on Twitter by learning to predict accuracy assessments in two credibility-focused Twitter datasets: CREDBANK, a crowdsourced dataset of accuracy assessments for events in Twitter, and PHEME, a dataset of potential rumors in Twitter and journalistic assessments of their accuracies.
We apply this method to Twitter content sourced from BuzzFeed's fake news dataset and show  models trained against crowdsourced workers outperform models based on journalists' assessment and models trained on a pooled dataset of both crowdsourced workers and journalists.
All three datasets, aligned into a uniform format, are also publicly available.
A feature analysis then identifies features that are most predictive for crowdsourced and journalistic accuracy assessments, results of which are consistent with prior work.
We close with a discussion contrasting accuracy and credibility and why models of non-experts outperform models of journalists for fake news detection in Twitter.
\end{abstract}


%
\IEEEpeerreviewmaketitle

\section{Introduction}

Measuring accuracy and credibility in text are well-studied topics in disciplines from psychology to journalism\cite{memon2003psychology,Maier2005,Fogg:1999:ECC:302979.303001}.
The proliferation of large-scale social media data and its increasing use as a primary news source  \cite{Liu2013}, however, is forcing a re-examination of these issues.
Past approaches that relied on journalistically trained ``gatekeepers'' to filter out low-quality content are no longer applicable as social media's volume has quickly overwhelmed our ability to control quality manually.
Instead, platforms like Twitter and Facebook have allowed questionable and inaccurate ``news'' content to reach wide audiences without review.
Social media users's bias toward believing what their friends share and what they read regardless of accuracy allows these fake stories to propagate widely through and across multiple platforms\cite{Mitra2015}.
Despite research into rumor propagation on Twitter \cite{Kwon2013,Starbird2014,Qazvinian:2011:RIM:2145432.2145602}, fake image sharing in disaster aftermath \cite{Gupta2013}, and politically motivated ``astroturfing'' \cite{DBLP:journals/corr/abs-1011-3768}, rumor and ``fake news'' are becoming increasingly problematic.
Computational methods have proven useful in similar contexts where data volumes overwhelm human analysis capabilities.
Furthermore, regularities in bot behavior \cite{Starbird2017} and financially motivated sensationalists \cite{Sydell2016} suggest machine learning-based approaches could help address these quality issues.

In this paper, we present a method for automating ``fake news'' detection in Twitter, one of the most popular online social media platforms.
This method uses a classification model to predict whether a thread of Twitter conversation will be labeled as accurate or inaccurate using features inspired by existing work on credibility of Twitter stories \cite{Castillo2013,Kwon2013}.
We demonstrate this approach's ability to identify fake news by evaluating it against the BuzzFeed dataset of 35 highly shared true and false political stories curated by Silverman et al. \cite{Silverman2016} and extracted from Twitter.
This work is complicated by the limited availability of data on what is ``fake news'' online, however, so to train this system, we leverage two Twitter datasets that study credibility in social media: the PHEME journalist-labeled dataset \cite{Zubiaga2015b} and the CREDBANK crowdsourced dataset \cite{Mitra2015}.
PHEME is a curated data set of conversation threads about rumors in Twitter replete with journalist annotations for truth, and CREDBANK is a large-scale set of Twitter conversations about events and corresponding crowdsourced accuracy assessments for each event.

Results show our accuracy prediction model correctly classifies two-thirds of the Twitter fake news stories and outperforms prior work in this area.
Furthermore, accuracy models generated from crowdsourced workers outperform models trained on journalists in classifying potentially fake Twitter threads.
Feature analysis also shows crowdsourced workers' accuracy assessments are more influenced by network effects while journalists' assessments rely more on tweet content and language.

This work makes the following contributions:
\begin{itemize}
\item An automated mechanism for classifying popular Twitter threads into true and fake news stories,
\item An analysis of the different features used by journalists and crowdsourced workers/non-experts in assessing accuracy in social media stories, and
\item An aligned collection of three datasets that capture accuracy judgements across true and false stories.
\end{itemize}

\section{Relevant Work and Datasets}

Social media's explosions in popularity has enabled research into credibility in the online context, especially on microblogging platforms.
Several previous efforts have proposed methods for evaluating the credibility of a given tweet \cite{Qazvinian:2011:RIM:2145432.2145602} or user \cite{Kang:2012:MTS:2166966.2166998} while others have focused more on the temporal dynamics of rumor propagation \cite{Kwon2013}.
Most relevant to our paper, however, is the 2013 Castillo et al. work, which provides a comprehensive examination of credibility features in Twitter \cite{Castillo2013}.
This study was built on an earlier investigation into Twitter usage during the 2010 Chile earthquake, where Twitter played a significant role both in coordination and misinformation \cite{Mendoza:2010:TUC:1964858.1964869}.
The later study developed a system for identifying newsworthy topics from Twitter and leveraged Amazon's Mechanical Turk (AMT) to generate labels for whether a topic was credible, similar to CREDBANK but at a smaller scale.
Castillo et al. developed a set of 68 features that included characteristics of messages, users, and topics as well as the propagation tree to classify topics as credible or not.
They found a subset of these features, containing fifteen topic-level features and one propagation tree feature, to be the best performing feature set, with a logistic regression model achieving an accuracy of $64\%$ for credibility classification.
Given general users have difficulty judging correct and accurate information in social media \cite{StanfordHistoryEducationGroup2016,Starbird2014}, however, crowdsourced credibility assessments like these should be treated with caution.
The investigation presented herein builds on this past work by evaluating whether crowdsourced workers (as used in both CREDBANK and Castillo et al.) are valid accuracy assessment sources.

\subsection{The PHEME Rumor Dataset}

The PHEME rumor scheme data set was developed by the University of Warwick in conjunction with Swissinfo, part of the Swiss Broadcasting Company \cite{Zubiaga2015b}.
Swissinfo journalists, working with researchers from Warwick, constructed the PHEME data set by following a set of major events on Twitter and identifying threads of conversation that were likely to contain or generate rumors.
A ``rumor'' in this context was defined as an unverified and relevant statement being circulated, and a rumor could later be confirmed as true, false, or left unconfirmed.

During each rumor selected in the PHEME dataset, journalists selected popular (i.e., highly retweeted) tweets extracted from Twitter's search API and labeled these tweets as rumor or non-rumor. 
This construction resulted in a set of 330 labeled rumorous source tweets across 140 stories.
For each tweet in this labeled set, the authors then extracted follow-up tweets that replied to the source tweet and recursively collected descendant tweets that responded to these replies.
This collection resulted in a tree of conversation threads of 4,512 additional descendant tweets.
Journalists from Swissinfo labeled source tweets for each of these threads as true, false, or unverified.
Once this curated set of labeled source tweets and their respective conversation threads were collected, the PHEME data set was then made available to crowdsourced annotators to identify characteristics of these conversation threads.
This crowdsourced task asked annotators to identify levels of support (does a tweet support, refute, ask for more information about, or comment on the source tweet), certainty (tweet author's degree of confidence in his/her support), and evidentiality (what sort of evidence does the tweet provide in supporting or refuting the source tweet) for each tweet in the conversation.
Past work found disagreement and refutation in threads to be predictive of accuracy \cite{Castillo2013}, and these annotations of whether a tweet supports or refutes the original tweet help quantify this disagreement, which we leverage later.

Of the 330 conversation trees in PHEME, 159 were labeled as true, 68 false, and 103 unverified.

\subsection{The CREDBANK Dataset}

In 2015, Mitra and Gilbert introduced CREDBANK, a large-scale crowdsourced data set of approximately 37 million of which were unique.
The data set covered 96 days starting in October of 2014, broken down into over 1,000 sets of event-related tweets, with each event assessed for accuracy by 30 annotators from AMT \cite{Mitra2015}.
CREDBANK was created by collecting tweets from Twitter's public sample stream, identifying topics within these tweets, and using human annotators to determine which topics were about events and which of these events contained accurate content.
Then, the systems used Twitter's search API to expand the set of tweets for each event.

CREDBANK's initial set of tweets from the 96-day capture period contained approximately one billion  tweets that were then filtered for spam and grouped into one-million-tweet windows.
Mitra and Gilbert used online topic modeling from Lau et al. \cite{Lau2012} to extract 50 topics (a topic here is a set of three tokens) from each window, creating a set of 46,850 candidate event-topic streams.
Each potential event-topic was then passed to 100 annotators on AMT and labeled as an event or non-event, yielding 1,049 event-related topics (the current version of CREDBANK contains 1,377 events).
These event-topics were then sent to 30 additional AMT users to determine the event-topic's accuracy.

This accuracy annotation task instructed users to assess ``the credibility level of the Event'' by reviewing relevant tweets on Twitter's website (see Figure 5 in Mitra and Gilbert \cite{Mitra2015}).
Annotators were then asked to provide an accuracy rating on a 5-point Likert scale of ``factuality'' (adapted from Sauri et al. \cite{Saur2009}) from $[-2, +2]$, where $-2$ represented ``Certainly Inaccurate'' and $+2$ was ``Certainly Accurate'' \cite{Mitra2015}.
Annotators were required to provide a justification for their choice as well.
These tweets, topics, event annotations, and accuracy annotations were published as the CREDBANK dataset.\footnote{Available online \url{http://compsocial.github.io/CREDBANK-data/}}
Data provided in CREDBANK includes the three-word topics extracted from Twitter's sample stream, each topic's event annotations, the resulting set of event-topics, a mapping of event-topics' relevant tweets, and a list of the AMT accuracy annotations for each event-topic.
One should note that CREDBANK does not contains binary labels of event accuracy but instead has a 30-element vector of accuracy labels.

In CREDBANK,  the vast majority ($>95\%$) of event accuracy annotations had a majority rating of ``Certainly Accurate'' \cite{Mitra2015}.
Only a single event had a majority label of inaccurate: the rumored death of Chris Callahan, the kicker from Baylor University's football team, during the 2015 Cotton Bowl (this rumorous event was clearly false as Callahan was tweeting about his supposed death after the game).
After presenting this tendency towards high ratings, Mitra and Gilbert thresholds for majority agreement and found that 76.54\% of events had more than 70\% agreement, and 2\% of events had 100\% agreement among annotators.
The authors then chose 70\% majority-agreement value as their threshold, and 23\% of events in which less than 70\% of annotators agreed were ``not perceived to be credible'' \cite{Mitra2015}.
This skew is consistent with Castillo et al. \cite{Castillo2013}, where authors had to remove the ``likely to be true'' label because crowdsourced workers labeled nearly all topics thusly.
We address this bias below.

\subsection{BuzzFeed News Fact-Checking Dataset}

In late September 2016, journalists from BuzzFeed News collected over 2,000 posts from nine large, verified Facebook pages (e.g., Politico, CNN, AddictingInfo.org, and Freedom Daily) \cite{Silverman2016}.
Three of these pages were from mainstream media sources, three were from left-leaning organizations, and three were from right-leaning organizations.
BuzzFeed journalists fact-checked each post, labeling it as ``mostly true,'' ``mostly false,'' ``mixture of true and false,'' or ``no factual content.''
Each post was then checked for engagement by collecting the number of shares, comments, and likes on the Facebook platform.
In total, this data set contained  2,282 posts, 1,145 from mainstream media, 666 from right-wing pages, and 471 from left-wing pages \cite{Silverman2016}.

\section{Methods}

This paper's central research question is whether we can automatically classify popular Twitter stories as either accurate or inaccurate (i.e., true or fake news).
Given the scarcity of data on true and false stories, however, we solve this classification problem by transferring credibility models trained on CREDBANK and PHEME to this fake news detection task in the BuzzFeed dataset.
To develop a model for classifying popular Twitter threads as accurate or inaccurate, we must first formalize four processes: featurizing accuracy prediction, aligning the three datasets, selecting which features to use, and evaluating the resulting models.

\subsection{Features for Predicting Accuracy}

Here, we describe 45 features we use for predicting accuracy that fall across four types: structural, user, content, and temporal.
Of these features, we include fourteen of the most important features found in Castillo et al., omitting the two features on most frequent web links.
Structural features capture Twitter-specific properties of the tweet stream, including tweet volume and activity distributions (e.g., proportions of retweets or media shares).
User features capture properties of tweet authors, such as interactions, account ages, friend/follower counts, and Twitter verified status.
Content features measure textual aspects of tweets, like polarity, subjectivity, and agreement.
Lastly, temporal features capture trends in the previous features over time, e.g., the slopes of the  number of tweets or average author age over time.
As mentioned, many features were inspired by or reused from Castillo et al. \cite{Castillo2013} (indicated by $^\star$).

\subsubsection{Structural Features}

Structural features are specific to each Twitter conversation thread and are calculated across the entire thread.
These features include the number of tweets, average tweet length$^\star$, thread lifetime (number of minutes between first and last tweet), and the depth of the conversation tree (inspired by other work that suggests deeper trees are indicators of contentious topics \cite{Tan:2016:WAI:2872427.2883081}).
We also include the frequency and ratio (as in Castillo et al.) of tweets that contain hashtags, media (images or video), mentions, retweets, and web links$^\star$.

\subsubsection{User Features}

While the previous set focuses on activities and thread characteristics, the following features are attributes of the users taking part in the conversations, their connectedness, and the density of interaction between these users.
User features include account age$^\star$; average follower-$^\star$, friend-, and authored status counts$^\star$; frequency of verified authors$^\star$, and whether the author of the first tweet in the thread is verified.
We also include the difference between when an account was created and the relevant tweet was authored (to capture bots or spam accounts).

This last user-centric feature, network density, is measured by first creating a graph representation of interactions between a conversation's constituent users.
Nodes in this graph represent users, and edges correspond to mentions and retweets between these users.
The intuition here is that highly dense networks of users are responding to each other's posts and endogenous phenomena.
Sparser interaction graphs suggest the conversation's topic is stimulated by exogenous influences outside the social network and are therefore more likely to be true.

\subsubsection{Content Features}

Content features are based on tweets' textual aspects and include polarity$^\star$ (the average positive or negative feelings expressed a tweet), subjectivity (a score of whether a tweet is objective or subjective), and disagreement$^\star$, as measured by the amount of tweets expressing disagreement in the conversation.
As mentioned in PHEME's description, tweet annotations include whether a tweet supports, refutes, comments on, or asks for information about the story presented in the source tweet.
These annotations directly support evaluating the hypothesis put forth in Mendoza, Poblete, and Castillo \cite{Mendoza:2010:TUC:1964858.1964869}, stating that rumors contain higher proportions of contradiction or refuting messages.
We therefore include these disagreement annotations (only a binary value for whether the tweet refutes the source).
Also borrowing from Castillo et al., we include the frequency and proportions of tweets that contain question marks, exclamation points, first/second/third-person pronouns, and smiling emoticons.

\subsubsection{Temporal Features}

Recent research has shown temporal dynamics are highly predictive when identifying rumors on social media \cite{Kwon2013}, so in addition to the frequency and ratio features described above, we also include features that describe how these values change over time.
These features are developed by accumulating the above features at each minute in the conversation's lifetime and converting the accumulated value to logarithmic space.
We then fit a linear regression model to these values in log space and use the slope of this regression as the feature's value, thereby capturing how these features increase or decrease over time.
We maintain these temporal features for account age, difference between account age and tweet publication time, author followers/friends/statuses, and the number of tweets per minute.

\subsection{Dataset Alignment}

While working with multiple datasets from different populations reduces bias in the final collection,  to compare the resulting models, we must translate these datasets into a consistent format.
That is, we must generate a consistent feature set and labels across all three datasets.

\subsubsection{Extracting Twitter Threads from BuzzFeed's Facebook Dataset}

The most glaring difference among our datasets is that BuzzFeed's data captures stories shared on Facebook, whereas CREDBANK and PHEME are Twitter-based.
Since Facebook data is not publicly available, and its reply structure differs from Twitter's (tweets can have arbitrarily deep replies whereas Facebook supports a maximum depth of two), we cannot compare these datasets directly.
Instead, we use the following intuition to extract Twitter threads that match the BuzzFeed dataset: Each element in the BuzzFeed data represents a story posted by an organization to its Facebook page, and all of these organizations have a presence on Twitter as well, so each story posted on Facebook is \textbf{also} shared on Twitter.
To align this data with PHEME and CREDBANK, we extract the ten most shared stories from left- and right-wing pages and search Twitter for these headlines (we use a balanced set from both political sides to avoid bias based on political leaning).
We then keep the top three most retweeted posts for each headline, resulting in 35 topics with journalist-provided labels, 15 of which are labeled ``mostly true,'' and 20 ``mostly false.''
Once these we identify these tweets, our BuzzFeed dataset mirrors the CREDBANK dataset in structure.

\subsubsection{Aligning Labels}

While the PHEME and BuzzFeed datasets contain discrete class labels describing whether a story is true or false (and points between), CREDBANK instead contains a collection of annotator accuracy assessments on a Likert scale.
We must therefore convert CREDBANK's accuracy assessment vectors into discrete labels comparable to those in the other datasets.
Given annotator bias towards ``certainly accurate'' assessments and the resulting negatively skewed distribution of average assessments, a labeling approach that addresses this bias is required.

Since majority votes are uninformative in CREDBANK, we instead compute the mean accuracy rating for each event, the quartiles across all mean ratings in CREDBANK, and construct discrete labels based on these quartiles.
First, the grand mean of CREDBANK's accuracy assessments is $1.7$, the median is $1.767$, and the 25th and 75th quartiles are 1.6 and 1.867 respectively.
In theory, events with mean ratings on the extreme ends of this spectrum should capture some latent quality measure, so events below or above the minimum or maximum quartile ratings in CREDBANK are more likely to be inaccurate and accurate respectively.
To construct ``truth'' labels from this data, we use the top and bottom $15\%$ quantiles, so events with average accuracy ratings more than $1.9$ become positive samples or less than $1.467$ become negative samples.
These quantiles were chosen (rather than quartiles) to construct a dataset of similar size to PHEME.
Events whose mean ratings are between these values are left unlabeled and removed from the dataset.
This labeling process results in 203 positive events and 185 negative events.


\subsubsection{Capturing Twitter's Threaded Structure}

Another major difference between PHEME and CREDBANK/BuzzFeed is the form of tweet sets: in PHEME, topics are organized into threads, starting with a popular tweet at the root and replies to this popular tweet as the children.
This threaded structure is not present in CREDBANK or our BuzzFeed data as CREDBANK contains all tweets that match the related event-topic's three-word topic query, and BuzzFeed contains popular tweeted headlines.
To capture thread depth, which may be a proxy for controversy \cite{Tan:2016:WAI:2872427.2883081}, we adapt CREDBANK's tweet sets and BuzzFeed's popular tweet headlines into threads using PHEME's  thread-capture tool.
For our BuzzFeed data, we use the popular headline tweets as the thread roots and capture replies to these roots to construct the thread structure mimicking PHEME's.
In CREDBANK, we identify the most retweeted tweet in each event and use this tweet as the thread root.
Any CREDBANK thread that has no reactions gets discarded, leaving a final total of 115 positive samples and 95 negative samples.

\subsubsection{Inferring Disagreement in Tweets}

One of the more important features suggested in Castillo et al. is the amount of disagreement or contradiction present in a conversation \cite{Castillo2013}.
PHEME already contained this information in the form of ``support'' labels for each reply to the thread's root, but CREDBANK and our BuzzFeed data lack these annotations.
To address this omission, we developed a classifier for identifying tweets that express disagreement.
This classifier used a combination of the support labels in PHEME and the ``disputed'' labels in the CreateDebate segment of the Internet Argument Corpus (IACv2) \cite{Abbott2015}.
We merged PHEME's support labels and IACv2 into a single ground-truth dataset to train this disagreement classifier.
Augmenting PHEME support labels with the IAC was necessary to achieve sufficient area under the receiver operating characteristic curve of $72.66\%$.

This disagreement classifier modeled tweet and forum text bags of unigrams and bigrams.
After experimenting with support vector machines, random forests, and naive Bayes classifiers, we found stochastic gradient descent to be the best predictor of disagreement and disputed labels.
10-fold cross validation of this classifier achieved a mean area under the receiver operating characteristic curve of $86.7\%$.
We then applied this classifier to the CREDBANK and BuzzFeed threads to assign disagreement labels for each tweet.
A human then reviewed a random sample of these labels.
While human annotators would be better for this task, an automated classifier was preferable given CREDBANK's size.

\subsection{Per-Set Feature Selection}

The previous sections present the features we use to capture structure and behavior in potentially false Twitter threads.
Our objective is to use these features to train models capable of predicting labels in the PHEME and CREDBANK datasets and evaluate how these models transfer to the BuzzFeed fake news dataset, but machine learning tasks are often sensitive to feature dimensionality.
That is, low-quality features can reduce overall model performance.
To address this concern, we perform a recursive feature elimination study within PHEME and CREDBANK to identify which features are the most predictive of accuracy in their respective datasets.

For each training dataset (i.e., CREDBANK and PHEME), we evaluate feature performance by measuring the area under the receiver operating characteristic curve (ROC-AUC) for a model trained using combinations of features.
The area under this ROC curve characterizes model performance on a scale of 0 to 1 (a random coin toss would achieve a ROC-AUC of 0.5 for a balanced set).
For each feature set, we perform thirty instances of 10-fold cross-validation using a 100-tree random forest classifier (an ensemble made of 100 separate decision trees trained on a random feature subset) to estimate the ROC-AUC for that feature set.

With the classifier and evaluation metric established, our feature selection process recursively removes the least performant feature in each iteration until only a single feature remains.
The least performant feature is determined using a leave-one-out strategy: in an iteration with $k$ features, $k$ models are evaluated such that each model uses all but one held-out feature, and the feature whose \emph{exclusion} results in the \emph{highest} ROC-AUC is removed from the feature set.
This method identifies which features hinder performance since removing important features will result in losses in ROC-AUC score, and removing unimportant or bad features will either increase ROC-AUC or have little impact.
Given $k$ features, the process will execute $k-1$ iterations, and each iteration will output the highest scoring model's ROC-AUC.
By inspecting these $k-1$ maximum scores, we determine the most important feature subset by identifying the iteration at which the maximum model performance begins to decrease.

\subsection{Evaluating Model Transfer}

Once the datasets are aligned and the most performance feature subsets in CREDBANK and PHEME are identified (these feature subsets are constructed separately and may not overlap), we can then evaluate how well each dataset predicts truth in the BuzzFeed dataset.
This evaluation is performed by restricting each source dataset (either CREDBANK or PHEME) to its most performant feature subset and training a 100-tree random forest classifier on each source.\footnote{We tested other classifiers here as well, and they all performed approximately equally.}
Each resulting classifier is applied to the BuzzFeed dataset, again restricted to the source dataset's most performant feature set, and the ROC-AUC for that classifier is calculated using the BuzzFeed journalists' truth labels.
This training and application process is repeated 20 times, and we calculate the average ROC-AUC across these repetitions.
We also build a third classification model by pooling both CREDBANK and PHEME datasets together and using the union of most performant features in each set.
We then plot the ROC curves for both source datasets, the pooled dataset, and a random baseline that predicts BuzzFeed labels through coin tosses and select the highest-scoring model.

\section{Results}

\subsection{Feature Selection}

\begin{figure}[htbp]
\begin{center}
\includegraphics[width=0.45\textwidth]{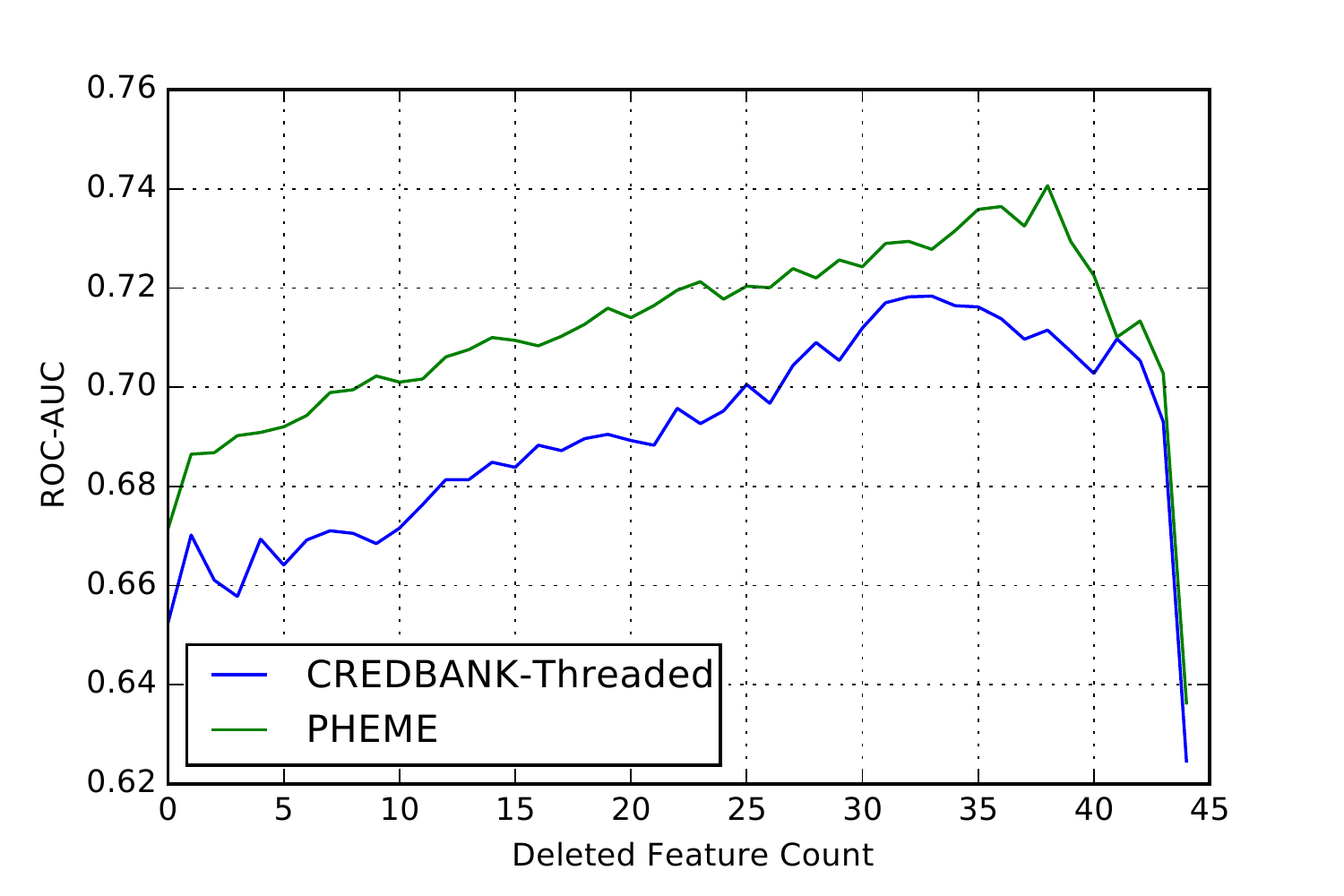}
\caption{Feature Elimination Study}
\label{fig:featureElimination}
\end{center}
\end{figure}

Recursively removing features from our models and evaluating classification results yielded significantly reduced feature sets for both PHEME and CREDBANK, the results of which are shown in Figure \ref{fig:featureElimination}.
The highest performing feature set for PHEME only contained seven of the 45 features: proportions and frequency of tweets sharing media; proportions of tweets sharing hashtags; proportions of tweets containing first- and third-person pronouns; proportions of tweets expressing disagreement; and the slope of the average number of authors' friends over time.
The top ten features also included account age, frequency of smile emoticons, and author friends.
This PHEME feature set achieved an ROC-AUC score of 0.7407 and correctly identified 66.93\% of potentially false threads within PHEME.

CREDBANK's most informative feature set used 12 of the 45 features: frequencies of smiling emoticons, tweets with mentions, and tweets with multiple exclamation or question marks; proportions of tweets with multiple exclamation marks, one or more question marks, tweets with hashtags, and tweets with media content; author account age relative to a tweet's creation date; average tweet length; author followers; and whether the a thread started with a verified author.
Proportions of tweets with question marks and multiple exclamation/question marks were not in the top ten features, however.
This feature set achieved an ROC-AUC score of 0.7184 and correctly identified 70.28\% of potential false threads within CREDBANK.

Of these feature subsets, only three features are shared by both crowdsourced worker and journalist models (frequency of smile emoticons and proportion of tweets with media or hashtags).
These results are also consistent with the difficulty in identifying potentially fallacious threads of conversation in Twitter discussed in Castillo et al. \cite{Castillo2013}.
Furthermore, both PHEME and CREDBANK's top ten features contain five of the 16 best features found in Castillo et al. \cite{Castillo2013}.
Despite these consistencies, our models outperform the model presented in this prior work (61.81\% accuracy in Castillo et al. versus 66.93\% and 70.28\% in PHEME and CREDBANK).
These increases are marginal, however, but are at least consistent with past results.

\subsection{Predicting BuzzFeed Fact-Checking}

\begin{figure}[htbp]
\begin{center}
\includegraphics[width=0.45\textwidth]{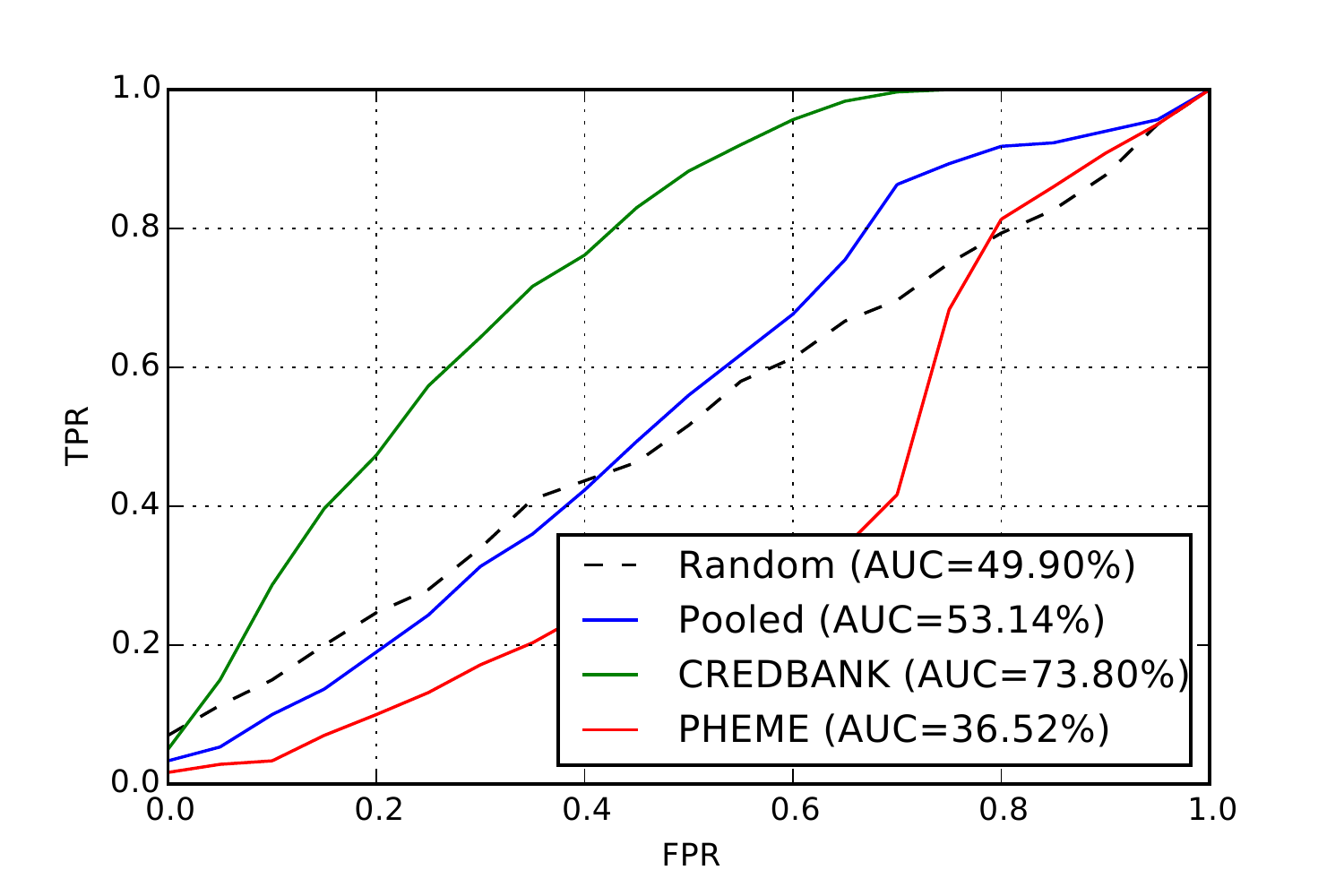}
\caption{Adapting to Fake News Classification}
\label{fig:toBuzzfeed}
\end{center}
\end{figure}

Applying the most performant CREDBANK and PHEME models to our BuzzFeed dataset shows  both the pooled and CREDBANK-based models outperform the random baseline, but the PHEME-only model performs substantially worse, as shown in Figure \ref{fig:toBuzzfeed}.
From this graph, CREDBANK-based models applied to the BuzzFeed performed nearly equivalently to performance in their native context, achieving a ROC-AUC of $73.80\%$ and accuracy of $65.29\%$.
The pooled model scores about evenly with the random baseline, with a ROC-AUC of $53.14\%$ and accuracy of $51.00\%$.
The PHEME-based model only achieved a ROC-AUC of $36.52\%$ and accuracy of $34.14\%$.
None of the dataset's results were statistically correlated with the underlying actual labels either,  with CREDBANK's $\chi^2(1, N=35)=2.803$, $p=0.09409$, PHEME's $\chi^2(1, N=35)=2.044$, $p=0.1528$, and the pooled model's $\chi^2(1, N=35)=0.2883$, $p=0.5913$. 

\section{Discussion}

Analysis of the above results suggest two significant results: First, models trained against non-expert, crowdsourced workers outperform models trained against journalists in classifying popular news stories on Twitter as true or fake.
Second, the limited predictive feature overlap in PHEME and CREDBANK suggest these populations evaluate accuracy in social media differently.

Regarding crowdsourced performance against the BuzzFeed dataset, since these stories were fact-checked by journalists, one might expect the PHEME model to perform better in this context.
We propose an alternate explanation: When a thread in Twitter starts with a story headline and link, the story's truth, as a journalist would define it, influences but does not dictate crowdsourced workers' perceptions of the thread.
Rather, it is this perception of accuracy that dictates how the story is shared.
Stated another way, the CREDBANK model captures user perceptions that drive engagement and sharing better than the PHEME model.
Furthermore, our CREDBANK model is more rooted in the Twitter context than PHEME since CREDANK assessors were asked to make their judgements based solely on the tweets they saw rather than the additional external information PHEME journalists could leverage.
While a CREDBANK assessor may have used external resources like search engines to check results, the majority of assessor justifications for their judgements were based on perception and how they felt rather than external fact checking \cite{Mitra2015}.
From this perspective, CREDBANK models may be more appropriate for a social media-based automated fake news detection task since both rely primarily on signals endogenous to social media (rather than external journalistic verification).
Finally, given the commensurate performance CREDBANK and PHEME exhibit in their native contexts, PHEME's poor performance for fake news suggest some fundamental difference between how endogenous rumors propagate in social media and how fake news is perceived and shared, but more work is needed here.

Along similar lines, though CREDBANK assessors are clearly biased towards believing what they read, our results show that the differences between story ratings capture some latent feature of accuracy.
That is, while users may be more likely to perceive false news stories as credible, their assessments suggest incorrect stories still receive lower scores.
Future research can use this information to correct for non-experts' bias towards believing what they read online, which may yield better models or better inform researchers about how fake news stories can be stopped before they spread.

Regarding contrasting accuracy models, we see diverging feature sets between PHEME and CREDBANK.
A review of the important features in each model suggest PHEME assessment is more linked to structural and content features rather than user, or temporal features.
CREDBANK assessments, on the other hand, focused more on different content markers, like formality of language (e.g., emoticons and many exclamation points), and user features, such as whether the tweet was from a verified author.
While both datasets are built on ``accuracy'' assessments, we theorize this question captures two separate qualities: for PHEME's journalists, ``accuracy'' is objective or factual truth, whereas CREDBANK's crowdsourced workers equate ``accuracy'' with credibility, or how believable the story seems. 
In PHEME, journalists evaluate the factual accuracy of conversation threads after ``a consensus had emerged about the facts relating to the event in question'' and after reviewing all the captured tweets relevant to that event \cite{Zubiaga2015b}.
CREDBANK assessors, as mentioned, focus more on perception of accuracy, or believability, in their justifications and are driven to make judgements rapidly by CREDBANK's ``real-time responsiveness'' \cite{Mitra2015}.
This distinction would also explain assessors' significant bias towards rating threads as accurate, which was present in both CREDBANK and Castillo et al. \cite{Castillo2013}, since readers are pre-disposed to believe online news \cite{Mackay2011,Starbird2014}.

Finally, by making an aligned version of this cross-platform dataset available, future research can explore differences between assessment populations.
Our results suggest journalists and crowdsourced workers use distinct signals in evaluating accuracy, which could be expanded and used to educate non-experts on which features they should focus when reading social media content.
Similarly, enhancing journalists' understanding of the features non-experts use when assessing accuracy may allow for better-crafted corrections to propagate through social media more rapidly.

\subsection{Limitations}

While the results discussed herein suggest crowdsourced workers provide a good source for identifying fake news, several limitations may influence our results.
This work's main limitation lies in the structural differences between CREDBANK and PHEME, which could affect model transfer.
If the underlying distributions that generated our samples diverge significantly, differences in feature sets or cross-context performance could be attributed to structural issues rather than actual model capabilities.
In future work, this limitation could be addressed by constructing a single data set of potential rumors and fake news threads and using both crowdsourced and journalist assessors to evaluate the same data.
This new data set would obviate any issues or biases introduced by the alignment procedure we employed herein.

Another potential limitation is this work's focus on popular Twitter threads.
We rely on identifying highly retweeted threads of conversation and use the features of these threads to classify stories, limiting this work's applicability only to the set of popular tweets.
Since the majority of tweets are rarely retweeted, this method therefore is only usable on a minority of Twitter conversation threads.
While a limitation, its severity is mitigated by the fact that fake news that is not being retweeted either is not gaining traction among the user base, or the user base has already identified it as fake.
Hence, our applicability to more popular tweets is valuable, as popular but fake stories have more potential to misinform than less popular fake stories.

\section{Conclusions}


This work demonstrates an automated system for detecting fake news in popular Twitter threads.
Furthermore, leveraging non-expert, crowdsourced workers rather than journalists provides a useful and less expensive means to classify true and false stories on Twitter rapidly.
Such a system could be valuable to social media users by augmenting and supporting their own credibility judgements, which would be a crucial boon given the known weaknesses users exhibit in these judgements.
These results may also be of value in studying propaganda on social media to determine whether such stories follow similar patterns.

\section*{Acknowledgements}

This research was supported by an appointment to the Intelligence Community Postdoctoral Research Fellowship Program at the University of Maryland, College Park, administered by Oak Ridge Institute for Science and Education through an interagency agreement between the U.S. Department of Energy and the Office of the Director of National Intelligence.

\section*{Data Availability}

Data and analyses are available online at: \url{https://github.com/cbuntain/CREDBANK-data}



\bibliographystyle{IEEEtran}
\bibliography{sources}

\end{document}